# Quantum Embedded Superstates

Nikita Nefedkin[1,2], Andrea Alú[1,3,4,*], and Alex Krasnok[1,*]

[1]*Photonics Initiative, Advanced Science Research Center, City University of New York, NY 10031, USA*

[2]*Moscow Institute of Physics and Technology, Moscow 141700, Russia*

[3]*Physics Program, Graduate Center, City University of New York, NY 10016, USA*

[4]*Department of Electrical Engineering, City College of The City University of New York, NY 10031, USA*

*To whom correspondence should be addressed: aalu@gc.cuny.edu, akrasnok@gc.cuny.edu

**Abstract**

Optical supercavity modes (superstates), i.e., hybrid modes emerging from the strong coupling of two nonorthogonal modes of an open cavity, can support ultranarrow lines in scattering spectra associated with quasi bound states in the continuum (quasi-BIC). These modes are of great interest for sensing applications as they enable compact systems with unprecedented sensitivity. However, these quasi-BIC sensors obey the shot-noise limit, which may be overcome only in quantum sensors. Here, we unveil that a three-level quantum system (e.g., atom, quantum dot, superconducting qubit) can be tailored to support the quantum analog of an embedded superstate with an unboundedly narrow emission line in the strong coupling regime. Remarkably, we demonstrate that the coupling of such a system with a cavity (e.g., plasmonic or dielectric nanoparticle, microcavity, microwave resonator) enables sensing properties with significantly reduced noise. Our results can be applied to a plethora of quantum platforms from microwave superconductors to cold atoms and quantum dots, opening interesting opportunities for quantum sensing and computing.

## Introduction

Elastic light scattering lies at the heart of the vast majority of sensors and detectors. Examples include tiny plasmonic nanosensors that enable detection down to the single-molecule level [1,2], and massive systems like LIGO used to detect gravitational waves [3]. Recent studies exploring unusual scattering phenomena, such as parity-time (PT) symmetry, exceptional points, topologically nontrivial phases, have unveiled sensing systems with exceptional properties [4]. One of the most interesting phenomena in this context is the so-called bound states in the continuum (BICs), also known as embedded eigenstates [5–10]. The idea of such eigenstates has been originally proposed by von Neumann and Wigner as a mathematical anomaly in quantum mechanics [11]. More recently, it has been generalized to different areas of wave physics, including acoustics, hydrodynamics, and photonics [5,10,12–18]. BICs are eigenstates with an unboundedly large Q-factor and correspondingly vanishing linewidth, corresponding to the coalescence of a scattering pole and zero at the same real frequency [4]. True BICs can arise only in structures either infinite in at least one dimension or employing lossless permittivity (permeability) materials with extreme values [5]. As a result, in realistic structures, BICs manifest themselves in the form of a narrow Fano resonance (quasi-BIC) with spectral width limited only by the proximity to the ideal requirements.

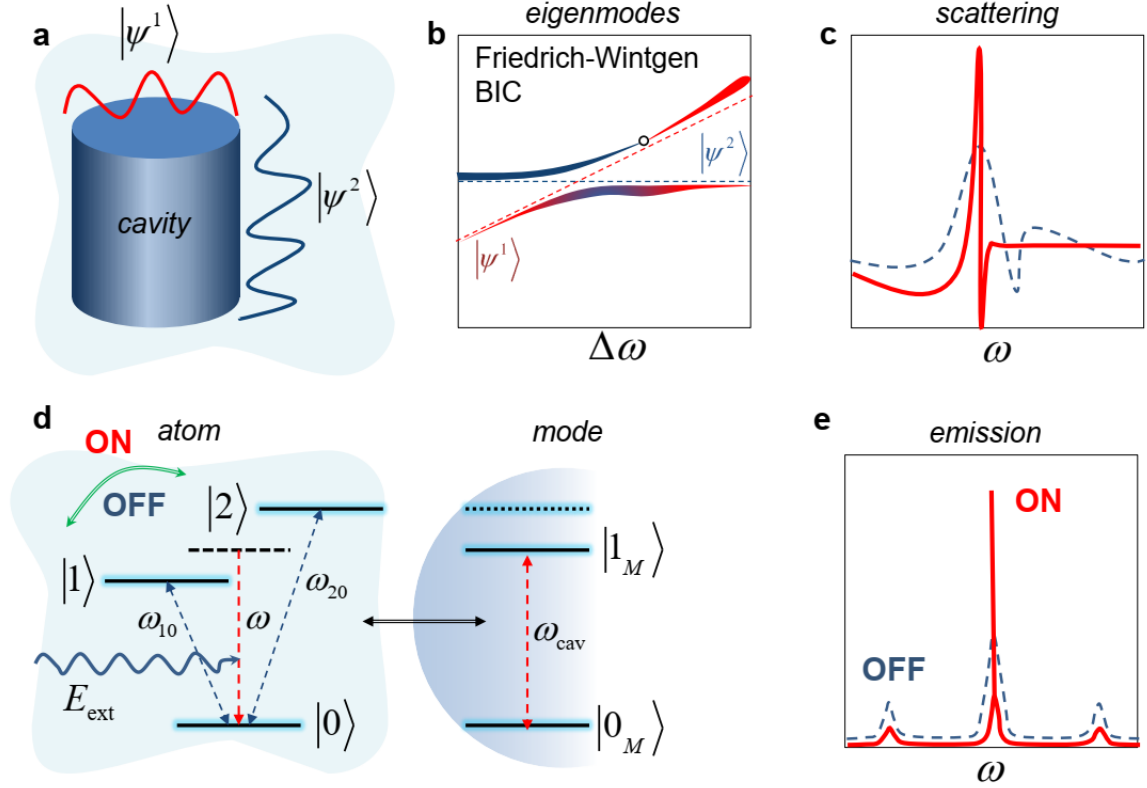

**Fig.1| Supercavity mode vs. quantum embedded superstate. a—c**, Illustration of an optical supercavity mode: a dielectric nano-cavity supports two strongly coupled nonorthogonal modes $|\psi^1\rangle$ and $|\psi^2\rangle$ (**a**). Hybridization of the bare modes leads to the formation of two branches of hybrid (dressed) modes. One of them can become dark (Friedrich-Wintgen BIC), while another one becomes brighter (**b**). In the scattering spectrum, such supercavity quasi-BIC state manifest

itself as a narrow Fano resonance (**c**). **d—e**, Illustration of a quantum embedded superstate. **d**, Sketch of a 3-level V-type quantum system (atom) interacting with a cavity mode, excited by an external monochromatic wave. A wisely tailored atom-field interaction leads to a narrow emission line and ultrahigh sensitivity (**e**).

Among various scenarios that support quasi-BICs, optical supercavity modes are of special interest for the purpose of this work. These modes emerge when an open cavity (Fig. 1a) supports two nonorthogonal modes $|\psi^1\rangle$ and $|\psi^2\rangle$ that hybridize into dressed states for a certain coupling strength, usually defined by the cavity geometry and material properties. With a suitable design, one of these states can become dark, giving rise to a Friedrich-Wintgen BIC [19,20], Fig. 1b, stemming from the destructive interference of the two modes. This dark supercavity mode manifests itself as a narrow Fano resonance (solid red curve in Fig. 1c) [21], of great interest for harmonic generation [22,23], quantum-entangled photons emission [24], nanolasers [25], and sensing, as demonstrated in a series of recent studies [14–16]. These quasi-BIC sensors are still bound to obey the classical shot-noise limit, motivating the exploration of analogous topics in the quantum limit for nonclassical states of light and matter.

The key phenomenon underlying the physics of quasi-BIC supercavity modes is the coherent destructive interference of two modes of a non-Hermitian cavity. In this work, we extend this concept to a fully quantum system and show that a three-level quantum system (e.g., atom, quantum dot, superconducting qubit) can support the *quantum analog of an embedded superstate* (QES) with unboundedly narrow emission line, Fig. 1d. This effect requires the simultaneous fulfillment of two conditions: the presence of so-called *quantum interference* (QI) between multiple atomic transition pathways (between the ground state $|0\rangle$ and excited states $|1\rangle$ and $|2\rangle$, Fig. 1d), and strong coupling of the excited states with an external field (freely propagating wave or standing cavity mode). The importance of QI in the system dynamics has been first underlined by Landau in his seminal paper on the density matrix approach [26]. The effects of QI can be introduced in a fully quantum treatment using the *generalized damping terms*, which are not generally included in semiclassical damping theories. The QI plays a vital role in a plethora of quantum phenomena, including quantum sensing [27], localization of atoms, spontaneous emission, induced transparency, bistability and gain without population inversion [28–33], and as we show in the following plays a crucial role in our proposed quantum sensors.

We explore QESs using the all-quantum Lindblad master equation formalism, where the presence of QI and the interaction with the reservoir degrees of freedom are rigorously taken into account. First, we show that a QES can be supported by a solitary standing quantum emitter (atom, QD). In [34], the crucial importance of QI on a V-type atom emission in the Bloch formalism has been explored in this context. Next, we demonstrate that the coupling of such a three-level V-type quantum system with a cavity (e.g., plasmonic or dielectric nanoparticle) provides unboundedly narrow emission spectra, Fig. 1, and it lifts the requirement for strong excitation enabling the operation in the *low-intensity quantum regime*, overcoming the conventional noise limitations of classical sensing systems.

## Results and discussion

**Theoretical description of a V-type atom interacting with a cavity.** We consider the case of a generalized three-level V-type atom (herein denoted as "atom"), interacting with a resonant cavity, see Fig. 1d. To analyze this system, we employ the Lindblad master equation formalism (see Methods and Supplementary Materials), which rigorously describes an extensive class of quantum systems (e.g., atoms, quantum dots, superconducting qubits, etc.) [35]. The Hamiltonian of such a system reads

$$\hat{H}_S = \hbar\omega_{cav}\hat{a}^+\hat{a} + \hbar\omega_{10}\hat{\sigma}_{10}^+\hat{\sigma}_{10} + \hbar\omega_{20}\hat{\sigma}_{20}^+\hat{\sigma}_{20} + \hat{V} \qquad (1)$$

where $\omega_{cav}$ is the frequency of the cavity mode, $\hat{a}$ and $\hat{a}^+$ are annihilation and creation operators of a quantum in the mode. The transition frequencies from the ground state $|0\rangle$ to states $|1\rangle$ and $|2\rangle$ are $\omega_{10}$ and $\omega_{20}$, respectively, Fig. 1d. The transition operators between the atom states are $\hat{\sigma}_{ij} = |j\rangle\langle i|$ and $\hat{\sigma}_{ij}^+ = |i\rangle\langle j|$, where $i, j = 0, 1, 2$. The interaction operator $\hat{V}$ accounts for the system interaction with the external electromagnetic field and subsystems with one another, which using the Jaynes-Cummings model for the cavity-mode interaction and the rotating wave approximation [35] reads

$$\hat{V} = \hbar\sum_k\left[\Omega_R^k\left(\hat{a}^+\hat{\sigma}_k + \hat{\sigma}_k^+\hat{a}\right) + \Omega_k\left(\hat{\sigma}_k + \hat{\sigma}_k^+\right)\right] + \hbar\Omega_a\left(\hat{a} + \hat{a}^+\right). \qquad (2)$$

The interaction constant of the atom and cavity is the Rabi constant $\Omega_R^i = -\mathbf{E}_{cav}\mathbf{d}_{ij}/\hbar$, where $\mathbf{d}_{ij} = \langle i | e\mathbf{r} | j \rangle$ is the matrix element of the dipole moment of the $i \to j$ transition. The variable $\mathbf{E}_{cav}$ is the electric near field of the cavity mode per one quantum, which can be found using the general relation $\frac{1}{8\pi}\int dV \frac{\partial(\text{Re}\varepsilon\omega)}{\partial \omega} |\mathbf{E}_{cav}|^2 = \hbar\omega_{cav}$ for a dispersive non-Hermitian system with complex permittivity $\varepsilon$ (see Refs. [36–39]). The constants $\Omega_k = -\mathbf{d}_k \mathbf{E}_{ext}\left[\exp(i\omega t) + \exp(-i\omega t)\right]/\hbar$ and $\Omega_a = -\mathbf{d}_{cav}\mathbf{E}_{ext}\left[\exp(i\omega t) + \exp(-i\omega t)\right]/\hbar$ correspond to the interaction of the atom $k$-th transition ($1 \leftrightarrow 0$, $2 \leftrightarrow 1$ and $2 \leftrightarrow 0$) and the cavity with the external field $\mathbf{E}_{ext}\left[\exp(i\omega t) + \exp(-i\omega t)\right]$, respectively.

In the rotating frame (see details in Methods), the system Hamiltonian (1) has the form

$$\hat{\tilde{H}}_S = \hbar\delta_{cav}\hat{a}^+\hat{a} + \hbar\delta_{10}\hat{\sigma}_{10}^+\hat{\sigma}_{10} + \hbar\delta_{20}\hat{\sigma}_{20}^+\hat{\sigma}_{20} + \hbar\Omega_R^{10}\left(\hat{a}^+\hat{\sigma}_{10} + \hat{\sigma}_{10}^+\hat{a}\right) + \\ \hbar\Omega_R^{20}\left(\hat{a}^+\hat{\sigma}_{20} + \hat{\sigma}_{20}^+\hat{a}\right) + \hbar\Omega_{10}\left(\hat{\sigma}_{10} + \hat{\sigma}_{10}^+\right) + \hbar\Omega_{20}\left(\hat{\sigma}_{20} + \hat{\sigma}_{20}^+\right) + \hbar\Omega_a\left(\hat{a} + \hat{a}^+\right), \quad (3)$$

where $\delta_{cav} = \omega_{cav} - \omega$ is the detuning between the cavity mode frequency and the frequency of the external wave; $\delta_{i0} = \omega_{i0} - \omega$ is the detuning of the $i$-th dipole transition frequency relative to the external wave frequency. The Hamiltonian (3) is Hermitian, as it describes the closed system. To take into account the relaxation processes, we should introduce the reservoir degrees of freedom, see Methods for details. The system interaction with reservoirs and interaction between subsystems is considered small in comparison with the cavity mode frequency and dipole transitions frequencies in the atom. After eliminating the reservoir degrees of freedom in the Born—Markov approximation [40], we arrive to the local master equation in Lindblad form [41,42], which describes the evolution of the system interacting with thermal baths

$$\frac{\partial}{\partial t}\hat{\rho} = -\frac{i}{\hbar}\left[\hat{\tilde{H}}_S, \hat{\rho}\right] + \mathcal{L}(\hat{\rho}), \quad (4)$$

where $\mathcal{L}(\hat{\rho}) = \frac{\gamma_a}{2}\mathcal{L}_1 + \frac{\gamma_1}{2}\mathcal{L}_2 + \frac{\gamma_2}{2}\mathcal{L}_3 + \frac{\Gamma}{2}\mathcal{L}_4 + \frac{\Gamma}{2}\mathcal{L}_5$, and $\mathcal{L}_1 = 2\hat{a}\hat{\rho}\hat{a}^+ - \hat{a}^+\hat{a}\hat{\rho} - \hat{\rho}\hat{a}^+\hat{a}$, $\mathcal{L}_2 = 2\hat{\sigma}_{10}\hat{\rho}\hat{\sigma}_{10}^+ - \hat{\sigma}_{10}^+\hat{\sigma}_{10}\hat{\rho} - \hat{\rho}\hat{\sigma}_{10}^+\hat{\sigma}_{10}$, $\mathcal{L}_3 = 2\hat{\sigma}_{20}\hat{\rho}\hat{\sigma}_{20}^+ - \hat{\sigma}_{20}^+\hat{\sigma}_{20}\hat{\rho} - \hat{\rho}\hat{\sigma}_{20}^+\hat{\sigma}_{20}$, $\mathcal{L}_4 = 2\hat{\sigma}_{10}\hat{\rho}\hat{\sigma}_{20}^+ - \hat{\sigma}_{20}^+\hat{\sigma}_{10}\hat{\rho} - \hat{\rho}\hat{\sigma}_{20}^+\hat{\sigma}_{10}$, $\mathcal{L}_5 = 2\hat{\sigma}_{20}\hat{\rho}\hat{\sigma}_{10}^+ - \hat{\sigma}_{10}^+\hat{\sigma}_{20}\hat{\rho} - \hat{\rho}\hat{\sigma}_{10}^+\hat{\sigma}_{20}$. Here, due to the fact that we opeate at optical frequencies, $\hbar\omega \gg kT$, we can assume that the stationary number of thermal quanta in the reservoir is negligible, i.e., $\bar{n} \ll 1$. In Eq. (4), $\gamma_a$ is the decay rate in the cavity mode, $\gamma_i$ is the spontaneous decay rate of the excited $i$ state of the atom. The terms proportional to $\Gamma$ are responsible for QI through the cross-coupling between dipole transitions $|1\rangle \leftrightarrow |0\rangle$ and $|2\rangle \leftrightarrow |0\rangle$. The QI effect is susceptible to the orientation of the dipole moments of the transitions, $\mathbf{d}_{10}$ and $\mathbf{d}_{20}$. If these dipole moments are parallel, then $\Gamma$ is maximum and it equals $\sqrt{\gamma_1\gamma_2}$ [34,43]. If they are orthogonal, the QI disappears, $\Gamma = 0$.

In our simulations, we assume that the interaction of the atom with the external field is insignificant due to the small dipole moment of the atom in comparison with the mode dipole moment, $\mathbf{d}_i / \mathbf{d}_{cav} \ll 1$, where $i = 10, 20$. Accordingly, we consider only the mode interaction with the external field. Thus, the external field excites the mode, which, due to large values of the Rabi constant (strong coupling regime), excites the atom.

**Quantum embedded superstate of a solitary atom.** We start by analyzing a solitary three-level V-type quantum system excited by an external monochromatic field. We obtain its Hamiltonian by excluding the cavity terms from Eq. (3): $\hat{\tilde{H}}_S^{at} = \hbar\delta_{10}\hat{\sigma}_{10}^+\hat{\sigma}_{10} + \hbar\delta_{20}\hat{\sigma}_{20}^+\hat{\sigma}_{20} + \hbar\Omega_{10}\left(\hat{\sigma}_{10} + \hat{\sigma}_{10}^+\right) + \hbar\Omega_{20}\left(\hat{\sigma}_{20} + \hat{\sigma}_{20}^+\right)$. In the symmetric case, when the decay rates in the atom are equal and the frequency of the external field falls precisely in the center between frequencies of the dipole transitions [$\gamma_1 = \gamma_2 = \gamma$ and $(\omega_{10} + \omega_{20})/2 = \omega$], the fluorescence can be quenched if the detunings meet the condition $\delta_{10}\Omega_{10}^2 + \delta_{20}\Omega_{20}^2 = 0$ and the dipole moments of the atom are parallel, $\Gamma = \sqrt{\gamma_1\gamma_2} = \gamma$ [34,43]. Let the energy distance between $|2\rangle$ and $|1\rangle$ be equal to $2\delta$, where $\delta = \omega_2 - \omega = \omega - \omega_1$, and $\Omega_{10} = \Omega_{20} = \Omega$. In these conditions, the eigenenergies and eigenstates read $E_0 = 0$, $E_\pm = \pm\hbar\sqrt{\delta^2 + 2\Omega^2}$, and $|\tilde{0}\rangle = -\frac{\Omega}{\tilde{\Omega}}|2\rangle + \frac{\Omega}{\tilde{\Omega}}|1\rangle + \frac{2\delta}{\tilde{\Omega}}|0\rangle$, $|\tilde{\pm}\rangle = \pm\frac{\tilde{\Omega}\pm\delta}{2\tilde{\Omega}}|2\rangle \pm \frac{\tilde{\Omega}\mp\delta}{2\tilde{\Omega}}|1\rangle + \frac{\Omega}{\tilde{\Omega}}|0\rangle$, respectively, where $\tilde{\Omega} = \sqrt{\delta^2 + 2\Omega^2}$. If the dipole moments $\mathbf{d}_{10}$ and $\mathbf{d}_{20}$ are parallel, the system is trapped in the "dark" eigenstate $|\tilde{0}\rangle$, and the system cannot emit [43].

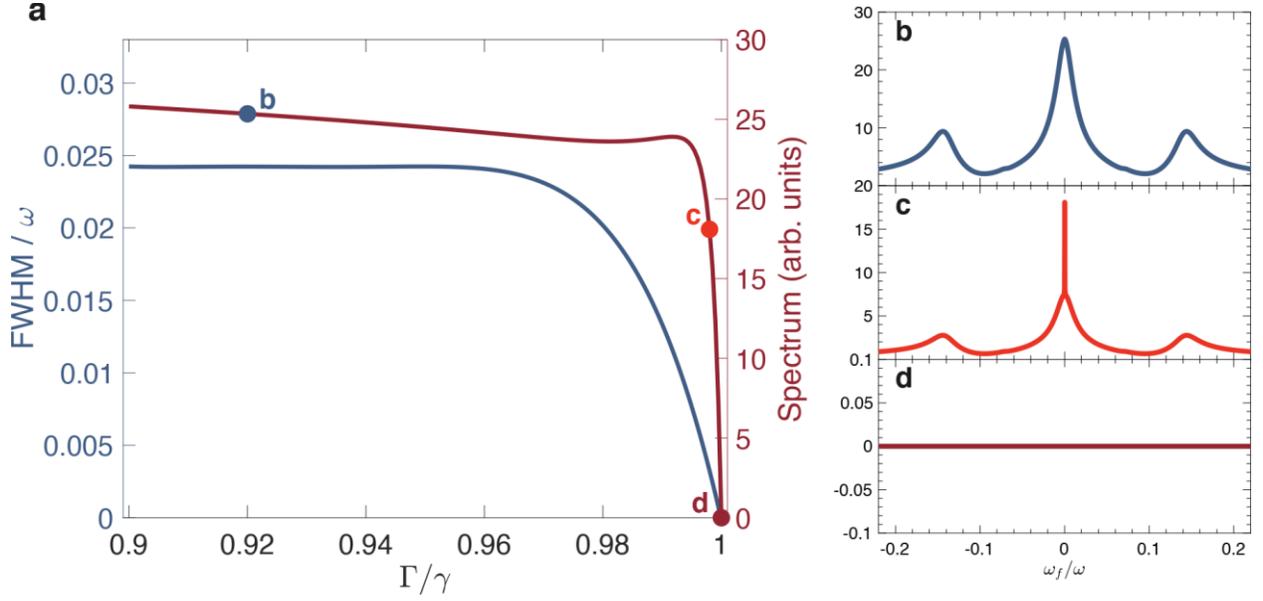

**Fig.2| Solitary three-level V-type atom in the external field. a,** Amplitude of the central peak of the resonance fluorescence spectrum of the atom as a function of $\Gamma/\gamma$ (red curve); full width at half maximum of the central peak versus $\Gamma/\gamma$ (blue curve). **b—c,** Fluorescence spectra for (**b**) $\Gamma/\gamma = 0.92$; (**c**) $\Gamma/\gamma = 0.998$ and (**d**) $\Gamma/\gamma = 1$ (quantum embedded superstate, QES). The other parameters: $\Omega = 5\gamma$, $\delta = \gamma/2$.

The fluorescence spectrum of the system can be calculated as (see Supplementary Materials)

$$S(\omega) = \Re \int_0^\infty \lim_{t\to\infty} \langle \hat{D}^+(t+\tau)\hat{D}(t)\rangle e^{-i\omega\tau} d\tau \tag{5}$$

where $\hat{D}(t) = \mathbf{d}_{10}\hat{\sigma}_{10} + \mathbf{d}_{20}\hat{\sigma}_{20} + \mathbf{d}_{cav}\hat{a}$ is the dipole moment of the atom and the cavity mode. We find the two-time averages $\langle \hat{D}^+(t+\tau)\hat{D}(t)\rangle$ using the quantum regression formula.

We simulated the fluorescence spectrum of a solitary atom for different values of $\Gamma$ in Fig. 2a. Both the amplitude and width of the spectral line do not change significantly until $\Gamma/\gamma \approx 0.98$, when the dipole moments of the $1\leftrightarrow 0$ and $2\leftrightarrow 0$ transitions become almost precisely parallel. Only when $\Gamma \cong \gamma$, the substantial suppression of fluorescence and narrowing of the spectral line can be observed. Figs. 2b-d demonstrate the fluorescence spectra in the three representative regimes denoted by points "b", "c", and "d" in Fig. 2a.

It is known that a conventional BIC manifests itself in the complex frequency plane as a scattering pole approaching the real frequency axis and canceling the corresponding scattering zero[44]. Now, we show that the discussed fluorescence suppression is closely connected with the imaginary part of one of the eigenfrequencies of the Lindblad matrix (see Supplementary Materials) approaching zero, Fig. 3. When the imaginary part of the central eigenfrequency in Fig. 3 becomes zero, the atom's population is trapped in the $|\tilde{0}\rangle$ eigenstate, and the fluorescence is suppressed.

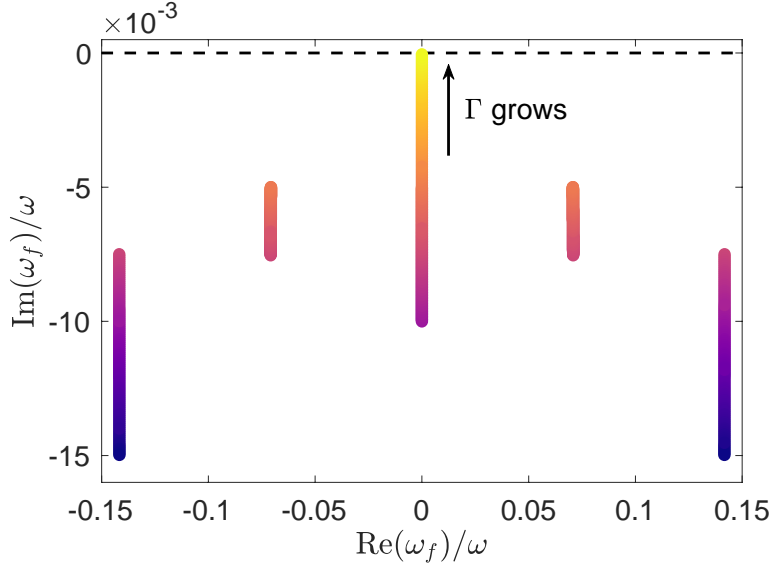

**Fig.3| Eigenfrequencies of the Lindblad operator.** Real and imaginary part of the eigenfrequencies of the Lindblad matrix. The real parts correspond to the following frequencies: $\pm 2\sqrt{\delta^2 + 2\Omega^2}$, $\pm\sqrt{\delta^2 + 2\Omega^2}$, $0$. The parameters are the same as in Fig. 2.

**Embedded superstate in an atom-cavity system.** It is important to note that this suppression is possible when the external field amplitude is very large, i.e., $\Omega \gg \gamma, \delta$. Although these fields are achievable in experiments, they are not convenient for the practical implementation of this effect. To overcome the limitation of large pumping fields in the solitary atom case, we place the atom near a resonant cavity. We study the variation of resonant fluorescence spectrum as a function of the cavity frequency and demonstrate very high sensitivity at low pump power under the condition that the dipole moments $\mathbf{d}_{10}$ and $\mathbf{d}_{20}$ are parallel.

The cavity mode is excited by an external optical wave, which is not in resonance with the cavity. We suppose that the pumping is small, such that the interaction constant between the external field and the mode is much less than the decay rates in the atom, $\Omega_a \ll \gamma_1, \gamma_2$, and is of the same order as the decay rate of the mode, $\Omega_a \sim \gamma_a$. Quenching follows $\delta_{10}(\Omega_R^{10})^2 + \delta_{20}(\Omega_R^{20})^2 = 0$, and we assume strong coupling regime between cavity and atom, $\Omega_R^i \gg \gamma_a, \gamma_1, \gamma_2$ and equal spontaneous decay rates of the dipole transitions, $\gamma_1 = \gamma_2 = \gamma$.

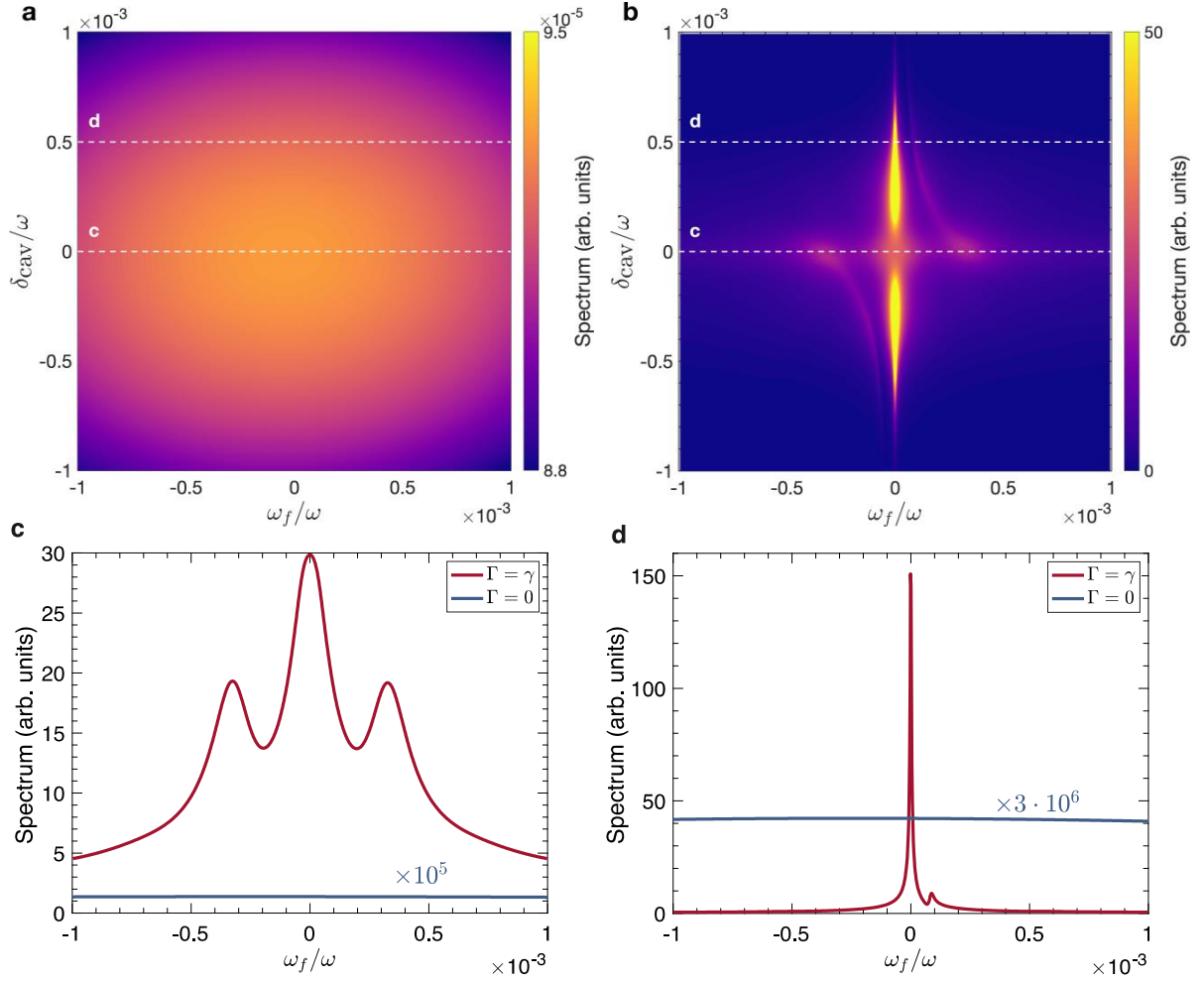

**Fig.4| Fluorescence spectra with and without quantum interference. a**, Spectrum of resonance fluorescence without quantum interference, $\Gamma = 0$, in a plane of fluorescence frequency, $\omega_f$, and the detuning between the frequency of the cavity mode and the external wave frequency, $\delta_{cav} = \omega_{cav} - \omega$. **b**, Spectrum with quantum interference, $\Gamma = \gamma$. **c,d**, Spectra profiles on the cross-sections along corresponding dashed lines. The other parameters: $\gamma_a / \gamma = 10^{-2}$, $|\omega_{20} - \omega_{10}| = 2\gamma$, $\Omega_R^{10} = \Omega_R^{20} = \Omega_R = 3\gamma$, $\Omega_a / \gamma = 2 \cdot 10^{-2}$.

Fig. 4 shows the fluorescence spectra in two limiting scenarios, $\Gamma = 0$ and $\Gamma = \gamma$, i.e., without (a) and with (b) quantum coherence. In Fig. 4a, the spectrum weakly depends on $\delta_{cav} = \omega_{cav} - \omega$ and almost does not change in the chosen frequency range for $\omega_f$, which is much smaller than $\Omega_R^i$.

The spectra profiles on the cross-sections along corresponding dashed lines are presented by blue curves in Figs. 4(c) and 4(d). At $\omega_{cav} = \omega$, the spectrum without QI has the form of a Mollow triplet with side peaks at $\sim \pm\sqrt{\delta^2 + 2\Omega_R^2}$ (see Supplementary Materials).

In remarkable contrast to the non-QI case, the QI scenario demonstrates an extremely narrow spectral line, Fig. 4b, revealing a strong dip at the resonance of the cavity mode and the external field. Although the dip in the center of Fig. 4b is profound, there is no total suppression of fluorescence, as in the case of a solitary atom, Fig. 2. This behavior is due to the contribution of the cavity to the emission and the finite dephasing rates. Out of resonance, the spectrum sharpens and its amplitude increases [Fig. 4d, red curve]. One of the side peaks disappears, whereas the second peak starts degenerating when $\omega_{cav}$ moves away from the resonance. Note that in this QI scenario, i.e., $\Gamma = \gamma$, the eigenfrequencies of the Lindblad operator with corresponding real parts approach the real axis (see Fig. S2 in Supplementary materials). Their behavior is similar to the case of the atom without a cavity, Fig. 3.

**Quantum embedded superstate for sensing applications.** We have shown how the QI significantly modifies the fluorescence spectra, sharpening them by three orders of magnitude. This effect significantly enhances the sensitivity to the cavity properties, ideally suited for quantum sensing applications. Here, we analyze the change in the fluorescence spectrum due to variations of the cavity frequency $\omega_{cav}$. This variation can be caused, for example, by a small permittivity change in the cavity [16,45–48]. Accordingly, we define the sensitivity of our QES-based system as the figure of merit

$$\text{FOM} = \frac{1}{\max(S)} \left| \frac{\partial S}{\partial \omega_{cav}} \right|, \qquad (6)$$

where $S$ is the fluorescence spectrum amplitude. Fig. 5 shows the calculated FOM of our system for the chosen parameters. We can see that the FOM is maximized near the spectrum peaks, as expected. At zero $\omega_f$, the spectrum profile has two distinguishable peaks, Fig. 5a. These peaks relate to the splitting of the mode levels due to the interaction with the external field. The spectral line here is very sharp, and it provides an excellent platform for sensor applications. The FOM reaches $\sim 5000$ along this cross-section. The intensity of the external wave can be comparatively small, which allows using this approach in low-intensity quantum applications.

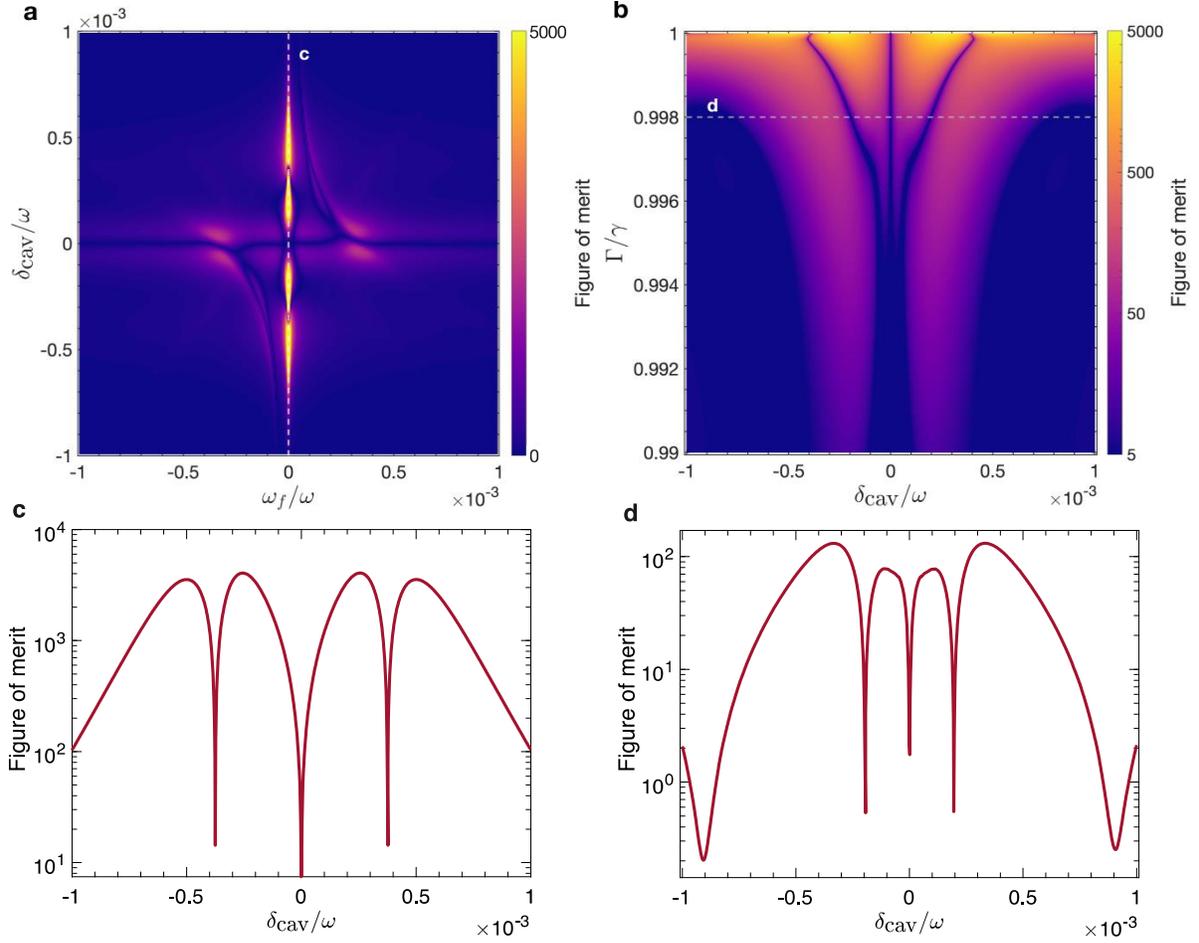

**Fig.5| Figure-of-merit of the QES-based quantum sensor.** Figure of merit in the plane of (**a**) $\omega_f$ and $\delta_{\text{cav}}$; (**b**) $\delta_{\text{cav}}$ and $\Gamma$ in logarithmic scale. **c,d,** Spectra profiles on the cross-sections along corresponding dashed lines in the logarithmic scale. The other parameters are the same as in Fig. 4.

Fig. 5b studies the influence on the FOM of atom dipole moments not being exactly parallel. Although the maximum sensitivity can be achieved at the parallel condition, satisfactory sensitivity with FOM $\sim 100$ is achievable up to $\Gamma = 0.99\gamma$. Thus, the effect of spectral line narrowing is pronounced only when the QI conditions are met. However, it stays robust for slight variations of the interference parameter.

Finally, we consider the statistical properties of the fluorescent emission in the parameter space formed by the frequency detuning between cavity mode and external wave frequency $\delta_{\text{cav}}$ and the collective dissipation rate $\Gamma$. Although the average number of quanta in the entire system is one, the resulting statistics of radiation can strongly differ from the single-photon regime. We study the second-order coherence function at zero time for the system radiation. As long as the dipole moment of the cavity mode is much larger than the dipole moments of the atom transitions, we can assume that almost the entire emission of the system originates from the mode. Thus, the coherence function has the form $g^{(2)}(0) = \langle \hat{a}^+ \hat{a}^+ \hat{a} \hat{a} \rangle / \langle \hat{a}^+ \hat{a} \rangle^2$ (see Refs. [35,49]). The deviation of the correlation function $g^{(2)}(0)$ from 1 gives the noise level in the system. Below, we show that quantum interference in our system brings the statistical properties of the interacting atom and cavity close to the properties of a solitary cavity mode, i.e., $g^{(2)}(0)$ tends to 2.

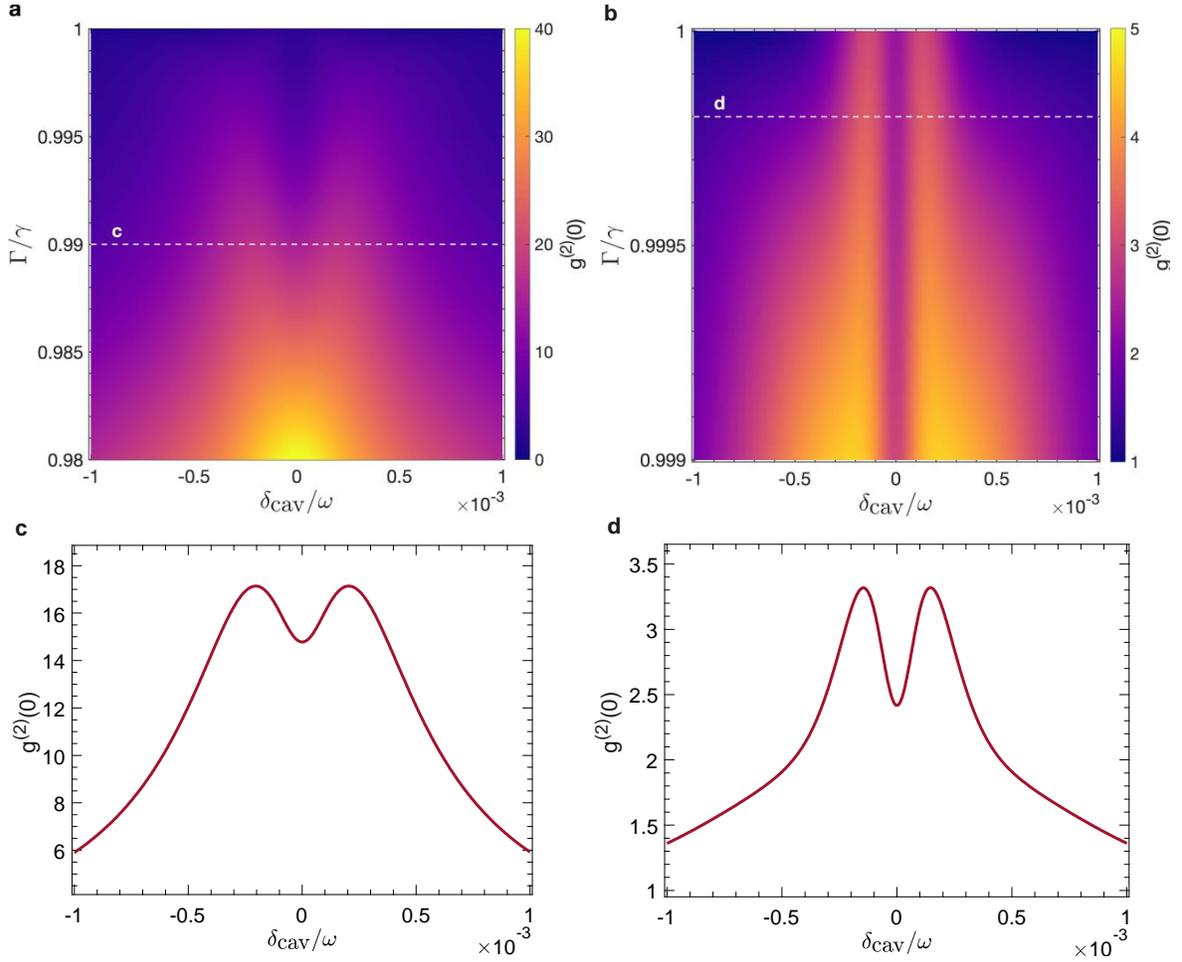

**Fig.6| Second-order coherence function of the EQS sensor. a**, Correlation function $g^{(2)}(0)$ in dependence on $\delta_{cav}/\omega$ and $\Gamma/\gamma$ with a zoomed area around full interference case (**b**). **c,d**, Cross sections denoted by white dashed lines. The other parameters are the same as in Fig. 4.

Fig. 6 shows the dependence of the second-order coherence function $g^{(2)}(0)$ on $\delta_{cav}$ and $\Gamma$. One can see substantial bunching of emitting photons with $g^{(2)}(0) \gg 10$ near resonance in the region of non-parallel dipole moments. Remarkably, the coherence function decreases considerably when approaching the condition of full interference ($\Gamma/\gamma=1$), see the zoomed area around $\Gamma/\gamma\approx 1$ in Fig. 6b. The coherence function has a two-peak shape as we vary the detuning. These peaks are connected with the frequency difference between the atom transitions. Exactly at resonance near the full interference, $g^{(2)}(0)$ has a significant dip with a value 2.4, see Fig. 6d. When moving away from resonance, $g^{(2)}(0)$ tends to 1 as the detuning grows. Nevertheless, in the full interference case, the value of $g^{(2)}(0)$ demonstrates the fastest convergence to 1. Thus, the system formed by a 3-levels atom strongly interacting with the cavity mode manifests bunching behavior in the whole parameter space in which the sensor operates. At resonance and for $\Gamma/\gamma\approx 1$, the $g^{(2)}(0)$ function takes its minimum value ~2, corresponding to the case in which either the first or the second dipole transition delivers a photon to the cavity mode, which then reradiates the photons with its usual statistics close to thermal emission [$g^{(2)}(0)=2$] [35]. This result shows that the effect of quantum interference and trapping in quantum embedded superstates significantly reduces the noise in the system, beyond the limits of classical sensors.

## Conclusion

In this work, utilizing an all-quantum Lindblad formalism, we have extended to a fully quantum scenario the concept of embedded superstates based on a three-level quantum system (e.g., atom, quantum dote, superconducting qubit). The key phenomenon beyond the demonstrated quantum supercavity modes is the quantum coherent destructive interference of two quantum states. These uniques quantum states are demonstrated to support unboundedly narrow emission lines with significantly improved second-order photon emission statistics. We have also shown that the coupling of such a three-level atom with an optical cavity (e.g., plasmonic or dielectric nanoparticle) paves the way to quantum sensing with remarkable performance. Namely, the proposed system provides a figure of merit of $5\cdot 10^3$ in the low-quantum regime of operation. The effect of quantum interference and traping in the quantum embedded superstate significantly reduces the noise in the system. This work unveils quantum embedded superstates, a novel state of quantum light-matter interaction, and suggest their use in advanced quantum-enhanced sensors with superior noise performance.

## Acknowledgements

This work was partially supported by the Simons Foundation, the National Science Foundation and the Department of Defense.

## Methods

**Lindblad master equation approach.** The system of the interacting atom and the cavity mode is an open quantum system, which interacts with reservoirs. We assume that the system is in a low-temperature environment and interacts only with the reservoirs of the EM modes of the free space. Thus, we consider the case when the energy of the system dissipates only through the radiation into free space. The Hamiltonian of the EM modes of free space and the interaction Hamiltonian of the system with the reservoir have the form

$$\hat{H}_R + \hat{H}_{SR} = \sum_{\mathbf{k},\lambda} \hbar \omega_\mathbf{k} \hat{a}^+_{\mathbf{k},\lambda} \hat{a}_{\mathbf{k},\lambda} + \sum_{\mathbf{k},\lambda} \hbar \omega_\mathbf{k} \hat{b}^+_{\mathbf{k},\lambda} \hat{b}_{\mathbf{k},\lambda} - \sum_{i,\mathbf{k},\lambda} \sqrt{\frac{2\pi \hbar \omega_\mathbf{k}}{V}} \mathbf{e}_{\mathbf{k},\lambda} \mathbf{d}_i \left( \hat{a}_{\mathbf{k},\lambda} + \hat{a}^+_{\mathbf{k},\lambda} \right) - \sum_{\mathbf{k},\lambda} \sqrt{\frac{2\pi \hbar \omega_\mathbf{k}}{V}} \mathbf{e}_{\mathbf{k},\lambda} \mathbf{d}_{cav} \left( \hat{b}_{\mathbf{k},\lambda} + \hat{b}^+_{\mathbf{k},\lambda} \right), \quad (M1)$$

where $i = 10, 20$. Note that we consider two uncorrelated reservoirs for the atom and the cavity. The system Hamiltonian (1) depends on time. Therefore, to get rid of this time dependence, we should perform the unitary transformation $\hat{U}(t) = \exp\left[ -i\hbar\omega \left( \hat{a}^+ \hat{a} + \hat{\sigma}^+_{10} \hat{\sigma}_{10} + \hat{\sigma}^+_{20} \hat{\sigma}_{20} \right) t \right]$ ($\omega$ is the frequency of the external field) of the Hamiltonian $\hat{H} + \hat{H}_R + \hat{H}_{SR}$ and move to the rotating frame:

$$\hat{\tilde{H}}_S = \hat{U}^+ \hat{H}_S \hat{U} - i\hat{U}^+ \frac{\partial \hat{U}}{\partial t} = \\ \hbar \delta_{cav} \hat{a}^+ \hat{a} + \hbar \delta_{10} \hat{\sigma}^+_{10} \hat{\sigma}_{10} + \hbar \delta_{20} \hat{\sigma}^+_{20} \hat{\sigma}_{20} + \hbar \Omega^{10}_R \left( \hat{a}^+ \hat{\sigma}_{10} + \hat{\sigma}^+_{10} \hat{a} \right) + \\ \hbar \Omega^{20}_R \left( \hat{a}^+ \hat{\sigma}_{20} + \hat{\sigma}^+_{20} \hat{a} \right) + \hbar \Omega_{10} \left( \hat{\sigma}_{10} + \hat{\sigma}^+_{10} \right) + \hbar \Omega_{20} \left( \hat{\sigma}_{20} + \hat{\sigma}^+_{20} \right) + \hbar \Omega_a \left( \hat{a} + \hat{a}^+ \right), \quad (M2)$$

$$\hat{\tilde{H}}_R = \hat{U}^+ \hat{H}_R \hat{U} = \sum_\nu \hbar \nu \hat{a}^+_{\nu,\lambda} \hat{a}_{\nu,\lambda} + \sum_{\nu'} \hbar \nu' \hat{b}^+_{\nu',\lambda} \hat{b}_{\nu',\lambda}, \quad (M3)$$

$$\hat{\tilde{H}}_{SR} = \hat{U}^+ \hat{H}_{SR} \hat{U} = \hbar \sum_v w_{1v} \left( \hat{\sigma}_{10}^+ e^{i\omega t} + \hat{\sigma}_{10} e^{-i\omega t} \right) \left( \hat{a}_{v,\lambda} + \hat{a}_{v,\lambda}^+ \right) +$$
$$\hbar \sum_v w_{2v} \left( \hat{\sigma}_{20}^+ e^{i\omega t} + \hat{\sigma}_{20} e^{-i\omega t} \right) \left( \hat{a}_{v,\lambda} + \hat{a}_{v,\lambda}^+ \right) + \quad (M4)$$
$$\hbar \sum_v w_{cav\,v'} \left( \hat{a}^+ e^{i\omega t} + \hat{a} e^{-i\omega t} \right) \left( \hat{b}_{v',\lambda} + \hat{b}_{v',\lambda}^+ \right).$$

In order to obtain the local Lindblad master equation in secular approximation, we assume that interaction between subsystems is much less than their frequencies $\Omega_R^i, \Omega_i \ll \omega_i$ and the coupling with reservoirs is small in comparison with Rabi frequencies[41,42]. Using the Born—Markov approximation and excluding the reservoir variables, we obtain the local Lindblad master equation [40,49]:

$$\frac{\partial}{\partial t} \hat{\rho} = -\frac{i}{\hbar} \left[ \hat{\tilde{H}}_S, \hat{\rho} \right] + \mathcal{L}(\hat{\rho}) \quad (M5)$$

$$\mathcal{L}(\hat{\rho}) = \frac{\gamma_a}{2} \left( 2\hat{a}\hat{\rho}\hat{a}^+ - \hat{a}^+\hat{a}\hat{\rho} - \hat{\rho}\hat{a}^+\hat{a} \right) +$$
$$\frac{\gamma_1}{2} \left( 2\hat{\sigma}_{10}\hat{\rho}\hat{\sigma}_{10}^+ - \hat{\sigma}_{10}^+\hat{\sigma}_{10}\hat{\rho} - \hat{\rho}\hat{\sigma}_{10}^+\hat{\sigma}_{10} \right) +$$
$$\frac{\gamma_2}{2} \left( 2\hat{\sigma}_{20}\hat{\rho}\hat{\sigma}_{20}^+ - \hat{\sigma}_{20}^+\hat{\sigma}_{20}\hat{\rho} - \hat{\rho}\hat{\sigma}_{20}^+\hat{\sigma}_{20} \right) + \quad (M6)$$
$$\frac{\Gamma}{2} \left( 2\hat{\sigma}_{10}\hat{\rho}\hat{\sigma}_{20}^+ - \hat{\sigma}_{20}^+\hat{\sigma}_{10}\hat{\rho} - \hat{\rho}\hat{\sigma}_{20}^+\hat{\sigma}_{10} \right) +$$
$$\frac{\Gamma}{2} \left( 2\hat{\sigma}_{20}\hat{\rho}\hat{\sigma}_{10}^+ - \hat{\sigma}_{10}^+\hat{\sigma}_{20}\hat{\rho} - \hat{\rho}\hat{\sigma}_{10}^+\hat{\sigma}_{20} \right).$$

The expressions for decay rates see, e.g., in Ref.[50]. In Eqs. (M5)-(M6), we assume that $T \ll \delta_{cav}, \delta_{10}, \delta_{20}$, such that $n(\omega_i) \ll 1$. Note that the last two terms are responsible for the quantum interference between $1 \leftrightarrow 0$ and $2 \leftrightarrow 0$ dipole transitions.